# Improving average ranking precision in user searches for biomedical research datasets

Douglas Teodoro[1,2], Luc Mottin[1,2], Julien Gobeill[1,2], Arnaud Gaudinat[2], Thérèse Vachon[3], Patrick Ruch[1,2]

[1]Text Mining Group, SIB Swiss Institute of Bioinformatics, 1227 Geneva, Switzerland
[2]Department of Information Science, HEG Geneva, HES-SO, 1227 Geneva, Switzerland
[3]Novartis Institutes for BioMedical Research – Text Mining Services (NIBR Informatics/TMS), Novartis Pharma AG, Postfach, 4002 Basel, Switzerland

## Abstract

Availability of research datasets is keystone for health and life science study reproducibility and scientific progress. Due to the heterogeneity and complexity of these data, a main challenge to be overcome by research data management systems is to provide users with the best answers for their search queries. In the context of the 2016 bioCADDIE Dataset Retrieval Challenge, we investigate a novel ranking pipeline to improve the search of datasets used in biomedical experiments. Our system comprises a query expansion model based on word embeddings, a similarity measure algorithm that takes into consideration the relevance of the query terms, and a dataset categorisation method that boosts the rank of datasets matching query constraints. The system was evaluated using a corpus with 800k datasets and 21 annotated user queries. Our system provides competitive results when compared to the other challenge participants. In the official run, it achieved the highest infAP among the participants, being +22.3% higher than the median infAP of the participant's best submissions. Overall, it is ranked at top 2 if an aggregated metric using the best official measures per participant is considered. The query expansion method showed positive impact on the system's performance increasing our baseline up to +5.0% and +3.4% for the infAP and infNDCG metrics, respectively. Our similarity measure algorithm seems to be robust, in particular compared to Divergence From Randomness framework, having smaller performance variations under different training conditions. Finally, the result categorization did not have significant impact on the system's performance. We believe that our solution could be used to enhance biomedical dataset management systems. The use of data driven expansion methods, such as those based on word embeddings, could be an alternative to the complexity of biomedical terminologies. Nevertheless, due to the limited size of the assessment set, further experiments need to be performed to draw conclusive results.

# Introduction

The use of search engines in the Web as content indexing but also as data providers, via e.g., hyperlinks and text snippets, such as in Google and PubMed, has changed the way digital libraries are managed from static catalogues and databases to dynamically changing collections in a highly distributed environment. Systems, such as PubMed, PubMed Central (PMC) and Europe PMC, successfully provide platforms for retrieving and accessing information in the scientific literature, an essential step for the progress of biomedical sciences. Biomedical research produces enormous amount of digital data, which is stored in a variety of formats and hosted in a multitude of different sites [1]. Datasets, such as individual-level genotype, protein sequence, pathology imaging and clinical trials, are a few common examples. To guarantee the quality of scientific research, and maximize societal investments, it is key that, in addition to manuscript content, supporting material and datasets used and produced are also accessible and easily searchable, enabling thus the reproduction of key research findings and, additionally, the generation of novel insights [2]. In particular, transparent and integrated access to these datasets is paramount for asserting reproducibility and reliability of research outcomes [2,3]. Consequently, more and more journals request authors to make datasets used and produced in their studies publicly available [4]. Furthermore, providing integrated access to research datasets fosters collaborations and speeds up scientific progress, and, with the advance in data analytics methods, allows the discovery of new knowledge from connected data [5].

Aware of the needs for data sharing in scientific research, several systems are being investigated and implemented to provide flexible and scalable information management that meets the scale and variety of data produced by the biomedical community [6]. For example, dbGaP provides public access to large-scale genetic and phenotypic datasets required for wide association study designs [7]. PhenDisco brings standardization of phenotype variables and of study metadata, and result ranking to dbGaP to improve search performance of phenotypes [8]. GigaDB not only hosts research datasets but also tools, such as executable workflows, and assigns a Digital Object Identifier to datasets, which can be then used and cited by other researchers [9]. OpenAIRE, a large-scale initiative funded by the European Commission, provides an open data infrastructure service that enables collection, interlink and access to research publications and datasets, and to projects of the European Commission and other national funding schemes [10]. Finally, the biomedical and healthCAre Data Discovery Index Ecosystem (bioCADDIE) consortium, funded by the US National Institute of Health Big Data to Knowledge program [11], aims at building a data discovery index that makes data findable, accessible, interoperable and reusable allowing thus biomedical researchers to more easily find, reanalyse, and reuse data [12]. A common characteristic of these systems is that they are all powered by an information retrieval engine that enables indexing of the dataset metadata and content, and allows end users to locate the appropriate research data from the set of indexed repositories.

In particular, bioCADDIE implements a search engine prototype, called DataMed, using ElasticSearch [13]. DataMed catalogues research datasets based on a standard core metadata model, called Data Tag Suite (DATS), which is designed to be generic, applicable to any type of dataset, and extendable to specialized data types [14] so that it can accommodate the diversity of research data types and formats. Research datasets are fed into the bioCADDIE architecture through an ingestion repository. Then, an indexing pipeline maps the disparate metadata from the diverse indexed repositories into the unified DATS model [13]. Due to the nature of bioCADDIE datasets and the system's intended use, DataMed faces several challenges to provide end users with a relevant ranking list of datasets for their queries. First, as largely studied in TREC challenges [15-17], in daily usage casual users tend to create small queries, varying usually from one to ten words in length. In general, top-ranked retrieval systems, such as those based on BM25-Okapi similarity measures, underperform in these scenarios in comparison with long queries [15]. Second, the heterogeneity of the research data corpora brings new challenges to search engines and similarity measure algorithms. Differently from the scientific literature, which is composed basically of text data types, research datasets are available in a myriad of multimodal formats, varying from gene expression and protein sequence data to results of bioassays and exhibiting contents generated over several years of development. Finally, constraints in query specifications posed by casual users, such as the dataset type of interest, makes the ranking task closer to the more sophisticated question-answering search and requires additional work in the original ranked list [18-21].

To improve search of biomedical research datasets, in this work we investigate a novel dataset ranking pipeline that goes beyond the use of metadata available in DataMed. In the context of the 2016 bioCADDIE Dataset Retrieval Challenge [22], which aimed at enhancing the indexing and searching features of bioCADDIE's DataMed, we have developed a few strategies to deal with short queries, disparate dataset corpora and user query constraints. Our approach includes a query expansion module that computes the word embedding for query terms and performs their expansion using the k-nearest embedded word vectors [23,24]. Query expansion based on similarity of embedded words has been successfully applied to enrich queries [25,26]. The word embedding algorithm is trained locally using different biomedical corpora. We have also developed an original similarity measure algorithm, named RTRL, which takes into account the relevance of the query terms to boost the ranking of datasets. The algorithm divides query terms into *non-relevant*, *relevant* and *key-relevant*. Then, it attempts to rank datasets containing most of key relevant terms on top. Finally, we have created a query and dataset classifier module that boosts datasets matching the query class in the ranking list [27,28]. The classifier uses the Universal Protein Resource[1] (UniProt) topics to constrain query and datasets to a set of biomedical topics [29]. Our system was evaluated using a corpus of around 800k datasets and 21 queries, and achieved competitive results in comparison with other participants. In the next sections, we describe in detail our approach and the results obtained.

---

[1] www.uniprot.org

# Materials and methods

In Figure 1, the architecture of our dataset information retrieval system is described: given as input the bioCADDIE dataset corpus and a set of user queries, we have (a) a pre-processing phase, where both the dataset corpus and the query are cleaned, and the query terms are enriched, (b) a ranking phase, where the query terms are compared to the dataset corpus, and a ranked list of datasets, which are likely answers to the query, is obtained, and (c) a post-processing phase, where the results are refined based on the categorization of the top-ranked datasets in relation to the input query. Processes (a), (b) and (c) are run online. In the background, we have three processes that run on a batch mode: (1) training of a word embedding model based on neural network, (2) bioCADDIE corpus indexing and (3) training of a dataset classification model. The word embedding model is used on the query pre-processing phase to expand query terms. The index created is used to actually provide the query answers, via a ranking model that computes the similarity between the indexed datasets and the query terms. Finally, the dataset classification model is used to constrain (or improve the ranking of) the results matching the input query constraints (e.g., gene expression dataset, protein sequence data, etc.).

## *Input data*

### bioCADDIE dataset corpus

In Table 1, an example of bioCADDIE's dataset representation following the DATS data model is presented. The DOCNO tag is a unique identifier for the dataset. The TITLE tag provides the dataset title, which is usually a meaningful and concise description of the dataset but that can be also sometimes just a code string identifying the dataset. The REPOSITORY tag indicates the repository hosting the original dataset content. The repository name is appended with a snapshot date, indicating the dataset version date. Finally, the METADATA tag is a JSON object with many attributes describing the dataset but also containing parts of the dataset itself. The METADATA tag is the most informative part of the dataset, containing altogether 96 attributes that vary from dataset description and organism to chemical formula. The first prototype of DataMed available at datamed.org integrates an initial set of 23 repositories covering 10 data types. In the challenge, a subset of this indexed collection was provided to the participants to build their information retrieval strategies. Table 2**Error! Reference source not found.** shows the repositories and the dataset distribution for the subset available. In total, there were 794,992 datasets distributed among 20 repositories. As we can see, there is a concentration of datasets in some repositories, in particular ClinicalTrials, BioProject, PDB and GEO, each of which contains more than 100k datasets. These top four repositories constitute 71.4% of the datasets and the top eight repositories constitute 99.5% of the total corpus.

### Training and assessment queries

Among others, bioCADDIE framework aims to answer user queries such as i) disease-based search across scales; ii) molecular-based search across organisms and scales, iii) molecular data/phenotype associations; and iv) behavioural and environmental data [13]. The challenge organizers provided 21 annotated queries covering these use cases. Six queries were made available at the beginning of the challenge to train the information retrieval system and 15 queries were used to do the official assessment. Table 3 shows some query examples as provided during the challenge. The organisers generated the initial query answers combining the top 1000 datasets retrieved by 4 information retrieval engines: Apache Lucene[2], Lemur Indri[3], Terrier[4], and Semantic Vectors[5]. Then, the documents retrieved were classified as not relevant, if they had less than 50% of the key query concepts; partially relevant, if they had all key query concepts but did not answer the question, or if they had the majority of the concepts but not all; and relevant, if they had all key query concepts and were an answer to the question. Key query concepts were defined according to the annotator's expertise. Finally, the query results were annotated by two annotators and corrected whenever necessary by a domain expert. For the test queries, a post-submission judgment was performed by pooling the participant's results and enriching the original set obtained with the 4 original information retrieval engines. The full query list is available in the *Train and test queries* section (Table S1) as supplementary material.

---

[2] https://lucene.apache.org/core
[3] https://www.lemurproject.org/indri
[4] http://terrier.org
[5] https://github.com/semanticvectors/semanticvectors

## *Pre-processing phase*

### Data treatment

In the pre-processing phase (Figure 1a), we treat the input data corpus and query. Greek letters are substituted by their literal names, stopwords are removed and terms are stemmed using Porter stemmer. Then, all non-alphanumeric characters are removed and terms are truncated to 20 characters. Finally, numerical sequences are replaced with the literal *_number_*.

### Query expansion

Recent results demonstrate the effectiveness of continuous space word embedding methods for analogy and word similarity tasks [25,26]. Continuous space embedding models project terms from a vocabulary into real-number vectors in a low-dimensional space [23]. Word representations in continuous space embeddings are modelled by assuming that if two words tend to occur in similar contexts, they are likely to play similar syntactic and semantic roles. We propose to use the *word2vec* algorithm [24] for word embedding although other continuous space methods behave similarly for query expansion [30]. In one of its model, so called continuous bag of words (CBOW), the word2vec algorithm implements a neural network that predicts a word given a small set of context words. Our query expansion model uses the k-nearest neighbours of the query terms in the embedding space. Formally, let $Q$ be a given user query consisting of the words $q_1, q_2, \ldots, q_m$. Let $w_1^i, w_2^i, \ldots, w_k^i$ be the k-nearest neighbours of $q_i$ in the embedding space. Then, the vectors $w_j^i$ constitute the set of term expansion candidates. We have created two expansion models based on this algorithm. The first model considers all query terms equally relevant for expansion. Original query terms are assigned to the unitary weight and the weights of the expanded terms are determined by their cosine similarity to the original query term in the embedding space. The second model implements a similar algorithm but with the difference that it reduces the weight of a query term and of its expansions by a factor *c* if the original term is not a *key-relevant* query term. We consider *key-relevant* terms as those with lower document frequency in the collection. For a given query, terms with document frequency lower or equal than 50% compared to the other query term document frequencies are considered *key-relevant* query terms. For example, in the query "transgenic mice", if the term *transgenic* has lower document frequency in the collection compared to the term *mice*, then *transgenic* will be considered a *key-relevant* query term and *mice* as a *relevant* query term. Thus, the weight of the term *mice* and of its expansions will be reduced by a loss factor *l*.

We use the Gensim word2vec library [31] to create the word embeddings with different input corpora: bioCADDIE (800k), PMC (36k) and Medline (200k). These collections were selected due to their relevance to biomedicine. It has been shown that locally trained word embeddings might provide superior word representation [25]. Therefore, we only considered external resources that have similar context to bioCADDIE's datasets. In particular, only articles annotated with UniProt terms were selected for the PMC and Medline collections. The word2vec neural network was trained using the continuous bag-of-words model with the dimensionality of the word embeddings set to 200 and the window size set to 5. The loss factor *l* for non *key-relevant* terms in the second query expansion model was empirically set to 1%.

## *Ranking phase*

For the dataset ranking (Figure 1b), we have investigated a similarity measure to robustly cope with the small query sizes expected from bioCADDIE users. As it has been shown in previous TREC competitions, the performance of more sophisticate ranking models, such as BM25-Okapi and Divergence from Randomness (DFR), tends to degrade when short queries are posed [15,32]. Simpler models often outperform these models. The main issue with more complex models for short queries is the tuning of the term frequency normalization parameter [32]. While for long queries this effect is smoothed due the abundance of context terms, for short queries the fine-tuning of the normalization parameters becomes much more relevant. In this context, we propose a simpler similarity model that we expect to be more robust to the changes in the training sets. Our model, called *robust term relevance logic* (RTRL), considers that a user query contains three types of terms: *non-relevant*, *relevant* and *key-relevant*. The rationale behind our similarity model is that if a document contains all *key-relevant* terms of a query, it should be ranked at the top. The remaining *relevant* terms are used to improve the ranking of these documents in the ranking list. For example, for the query *"Find data on T-cell homeostasis related to multiple sclerosis across all databases"*, after removing *non-relevant* query terms (i.e., query stopwords), we remain with the terms *T-cell*, *homeostasis*, *multiple* and *sclerosis*. Let us consider that the terms *T-cell* and *multiple* are relevant terms, and the terms *homeostasis* and *sclerosis* are the key-relevant terms for the query. Hence, documents containing the terms *homeostasis* and *sclerosis* should be at least as relevant to the query as documents containing *T-cell*, *homeostasis*, and *multiple* terms or as documents containing *T-cell*, *multiple* and *sclerosis* terms. More formally, for a query $Q = q_1, q_2, \ldots, q_n$ the weight of a key-relevant term is given by the equation

$$kw'' \geq (k-1)w'' + (L-k)w', \tag{1}$$

where $L$ is the number of query terms, $k$ is the number of key-relevant query terms, and $w'$ and $w''$ are the weights of the *relevant* and *key-relevant* query terms respectively. Additionally, to take into account the relevance of a query term to a document, the weight parameter of a term can be expanded into two components, $w = w_d + w_q$, where the component $w_d$ captures the relevance of a query term for a document and the component $w_q$ captures the term relevance within the query. Substituting this expansion in Eq. 1 and considering the worst case scenario, where top ranked documents contain only key-relevant terms with low document relevance, we have

$$k(w_d' + w_q'') \geq (k-1)(w_d'' + w_q'') + (L-k)(w_d'' + w_q'), \tag{2}$$

where $w_d'$ and $w_d''$ are the weights of a term with low and high document relevance respectively, and $w_q'$ and $w_q''$ are the weights of a term with low and high query relevance respectively (i.e., relevant and key-relevant terms). To solve Eq. 2, we can arbitrarily assign values to the weight variables $w'$ and $w''$ in such a way that the inequality constraint is respected. In our experiments, we fixed the values for the low document and query relevance weights to $w_d' = w_q' = 1$ and considered the weight of a high relevant document term twice the weight of a low relevant document term, i.e., $w_d'' \equiv 2w_d'$. Furthermore, we assume that the number of key and non-key relevant is roughly the same, i.e., $k \approx L/2$. Thus, we end up with a single free variable, the weight of a key-relevant term $w_q''$, which is a function of the query length and can be given by

$$w_q'' = c(2L - 2), \tag{3}$$

where $c \geq 1$ determines the gain in the *key-relevant* terms in relation to the other query terms. Then, the score of a document $d \in D$ will be given by the equation

$$\begin{aligned} \text{score}(d, Q) &= \sum_{t \in Q} w(t, d) \\ &= \sum_{t \in Q} (w_d + w_q) \end{aligned} \tag{4}$$

In our experiments, we use the same definition for relevant and key-relevant as presented in the *Query expansion* section, i.e., *key-relevant* terms have a lower collection occurrence $f_{t,D}$ when compared to *relevant terms*. For the document relevance, we consider high document relevance when the term document occurrence $f_{t,d}$ is more than one. Otherwise, the document relevance is low. Hence,

$$w_d = \begin{cases} w_d' = 1 & \text{if } f_{t,d} = 1 \\ w_d'' = 2w_d' & \text{if } f_{t,d} > 1 \end{cases}, \tag{5}$$

and

$$w_q = \begin{cases} w_q' = 1 & \text{if } f_{t,D} > \text{median}(f_{Q,D}) \\ w_q'' = c(2L-2) & \text{if } f_{t,D} \leq \text{median}(f_{Q,D}) \end{cases}. \tag{6}$$

*Non-relevant* terms are basically stopwords and are neglected during the similarity computation.

When applying the weighting model described in Eq. 4 to the query terms, documents will be ranked into relevance bins. For example, for a query with 4 terms, being $q_1$ and $q_2$ relevant terms and $q_3$ and $q_4$ key-relevant, if a document $d_1$ contains only one occurrence of $q_3$ and $d_2$ only one occurrence of $q_4$, and none of the other terms, then using Eq. 4, Eq. 5 and Eq. 6 these documents will be ranked in the same ranking bin ($score_{d1} = score_{d2} = 1$). However, the discrimination power of q₃ and q₄ might be different. Then, to rank these documents within their bin, we add a $tf \cdot idf$ component to Eq. 4, i.e.,

$$\text{score}(d, Q)_b = \sum_{t \in Q}(w_d + w_q)_b + tf \cdot idf_b, \quad (7)$$

where $0 < tf \cdot idf_b \leq 1$ and is computed for each score bin $\text{score}(d, Q)_b$. The $tf$ and $idf$ components are defined as:

$$tf(t, d) = 0.5 + 0.5 \frac{f_{t,d}}{\max\{f_{t',d} : t' \in d\}}, \quad (8)$$

$$idf(t, D) = \log \frac{|D|}{|\{d \in D : t \in d\}|}. \quad (9)$$

## *Post-processing phase*

In the post-processing phase (Figure 1c), the initial results obtained during the ranking phase are modified so that the rank of documents matching the query constraints is boosted. For example, for queries searching specifically for *gene expression* datasets, the datasets retrieved containing gene expression information shall have higher ranks than those that do not contain gene expression information, regardless of the appearance of the terms *gene* and *expression* in the dataset body. Our query and result categorization model is based on the UniProt categories. UniProt has been developing a categorization model for many years now, where UniProt curators classify and annotate biomedical datasets into 11 categories: *Expression*, *Family & Domains*, *Function*, *Interaction*, *Names*, *Pathology & Biotech*, *PTM/processing*, *Sequences*, *Structure*, *Subcellular location*, *Unclassified* (or miscellaneous classes). UniProt curators use these categories to annotate scientific articles available in PubMed. Each annotation provides a category, such as *Sequences*, and a qualifier to justify the citation, e.g., cited for 'NUCLEOTIDE SEQUENCE [MRNA]'. These classes cover relatively well the domain of bioCADDIE datasets and queries. A dataset class that is largely present in bioCADDIE but is not in UniProt is *Clinical trial*. Thus, we add this class to the other 11 UniProt classes to increase our classification coverage. We perform the re-ranking for the datasets originally ranked in the top 50% positions having the highest rank as reference. The datasets matching the query constraint classes are then multiplied by a gain factor *g*, which was set empirically to 10% for the official results.

## Query classification model

The query categorization is performed using simple string matching algorithm. 3-skip-2-grams, 2-skip-2-grams (2,2) and 1-gram are extracted from the query and matched against the UniProt categories and qualifiers. For the Clinical trial class, we defined a set of clinical trial keywords, namely *inclusion*, *exclusion*, *criteria*, *patients*, *subjects, stage, duration* and *study*. Then, we search for these keywords in the input query and classify them into the *Clinical trial* class in case a match is found.

## Dataset classification model

We use a multilayer perceptron (MLP) classifier to classify automatically bioCADDIE datasets against the UniProt classes mentioned above [29]. The input layer of the MLP is fed by documents (datasets) and the output layer is mapped to the UniProt classes (i.e., a multi-label multi-class classification task). The input documents (or datasets) are embedded into a vector space using the *paragraph2vec* model [33]. Paragraph2vec is a variation of the word2vec model that takes into account the document identifier in the context. The classifier is trained using 200k PubMed abstracts annotated with UniProt classes. We use the paragraph2vec implementation from Gensim (doc2vec) [31] to embed the documents/datasets. The classification model achieves a F1 score of 0.80. For the *Clinical trial* class, we use the keywords *clinical* and *trial* and *clinicaltrial* to categorize a document as clinical trial. If a bioCADDIE dataset contains one of these keywords they are classified into the *Clinical trial* class. In the online phase, new unclassified documents coming from the result set produced in the ranking phase are embedded in the paragraph2vec space and are fed into the MLP classifier, which produces the UniProt classes. In parallel, a string matching algorithm searches for the *Clinical trial* class keywords into the dataset content.

## Evaluation criteria

Performance results are reported using the inferred Average Precision (infAP) and inferred Normalized Discounted Cumulative Gain (infNDGC) metrics [34]. Both measures are designed to deal with incomplete judgements, as it is often the case for very large corpora, such as the collection indexed by DataMed. Hence, infAP and infNDCG deploy a randomly sampling approach to infer the mean AP (MAP) and NDCG performances, respectively. Additionally, we report the precision at 10 (P@10) considering partially relevant answers as relevant (+partial) but also as not relevant (-partial). These metrics compose the official measures of the 2016 bioCADDIE Dataset Retrieval Challenge. The organisers associated weight 2 to relevant, weight 1 to partially relevant, weight 0 to not relevant and weight -1 to unjudged documents in the gold standard query relevance file (see [35] for detailed information). We provide the results of the training and official assessments. Furthermore, we performed a 5-fold cross-validation using the post-judgement official gold standard. We compare the performance of our methods to a baseline approach based on the DFR model implemented by Terrier 4.1 [36,37]. Statistic hypothesis testing is performed using paired *t*-test with a two-tailed distribution and results with *P*-value smaller than .05 are considered statistically significant.

# Results

We assessed 5 information retrieval models using the bioCADDIE corpus and the annotated queries. The first model, *sibtex-1*, is a baseline model based on the DFR algorithm. For this model, in the pre-processing phase, the query and input corpus are treated and the similarity is computed using Terrier. The other 4 models use the methods described in the previous section. For the second model, *sibtex-2*, in the pre-processing phase, in addition to the query and corpus treatment, the query terms are expanded using the word2vec algorithm trained on the bioCADDIE and PMC collections, and in the post-processing phase the results are categorized using our classifier. The third model, *sibtex-3*, implements a similar algorithm to sibtex-2. The difference is that the weight of expanded terms derived from non key-relevant terms are reduced. For both models sibtex-2 and sibtex-3 the similarity score is obtained using Terrier's DFR algorithm. The fourth model, *sibtex-4*, uses the RTRL similarity model, which takes into account the relevance of the query terms in the ranking phase, and in the post-processing phase the results are categorized. Finally, the fifth model, *sibtex-5*, combines linearly the results of ranking model sibtex-3 and sibtex-4, i.e., sibtex-5 = $\alpha$sibtex-3 + $(1-\alpha)$sibtex-4, where $\alpha$ = 0.5 for the official submission. In resume, we assess the following method and model combinations:

- sibtex-1: DFR-based;
- sibtex-2: sibtex-1 + query expansion + results categorisation;
- sibtex-3: sibtex-1 + query expansion with penalized terms + results categorisation;
- sibtex-4: RTRL similarity measure + results categorisation;
- sibtex-5: linear combination of sibtex-3 and sibtex-4.

As an example of how the data processing works during an online search, let us consider the input query *"T1 – Find protein sequencing data related to bacterial chemotaxis across all databases"* submitted to model sibtex-2 (see *Train and test queries* section, provided as supplementary material, for the whole list of queries). In the pre-processing phase (Figure 1a), the cleaning process removes the non-relevant query terms (query stopwords) and reduces the remaining terms to their stem, leaving only the terms *"protein"*, *"sequenc"*, *"bacteri"* and *"chemotaxi"*. These terms are then expanded, resulting in an array with weights proportionally to their similarity to the original term. For example, the expansion of the term *chemotaxis* with k=3 results into the following weighted array: [chemotactic^0.089, chemoattractant^0.083, motility^0.080]. These terms are then stemmed and, together with the other original (with unitary weight) and expanded terms, processed by the ranking algorithm (Figure 1b). The ranking algorithm computes the query-to-dataset similarity and provides the ranked list of most similar datasets to the expanded query. Finally, in the post-processing phase (Figure 1c), the original query is classified into the UniProt/clinical trial classes using string matching against the UniProt class descriptors or clinical trial keywords. In this example, the query *T1* is classified as searching for *"Sequences"* datasets since the bigram "protein sequence" matches one of the UniProt class qualifiers for the *Sequences* class. In parallel, the result set containing the dataset retrieved is classified using the UniProt classifier. For example, the content of the dataset with identifier 719124, which appears in the top 1 for this query, is submit to the classifier and as output the classifier categorizes it as belonging to the *Sequences* class. Thus, its ranking score is boosted by 10%. This process is repeated for the top ranked datasets (we use a 50% threshold) and the final result set is obtained by rearranging the new dataset scores.

## *Performance results on the training set*

Table 4 shows the performance of 5 information retrieval models for the 6 training queries. As we can see, the model sibtex-3 outperforms the other models for the infAP and infNDCG metrics, with a marginal improvement over the baseline of +0.9% for the infAP metric and +0.4% for the infNDCG metric. For the P@10 metric, the models sibtex-4 and sibtex-5 outperform the other models, improving the baseline by +20%, with the latter having enhanced infAP and infNDCG outcomes with respect to the first. The optimized parameters for the different models are shown as supplementary material in Table S2 of the *Training parameters* section.

## *Performance results on the official test set*

The official performance results of the bioCADDIE challenge for our 5 models are displayed in Table 5. As we can notice, the sibtex-5 model outperforms all the other models for all metrics apart from infNDCG, for which the sibtex-4 model achieves the best performance. In particular, the sibtex-5 model outperforms the inferred average precision of the baseline by +21.89%. Apart from model sibtex-2, the other 3 models outperform the baseline for the infAP metric. There was a decrease in performance for the models sibtex-2 and sibtex-3 for the metric infNDCG with respect to the baseline, suggesting that the use of query expansion have a negative impact for this metric. The models using our similarity measure were able though to outperform the baseline for this metric. Specifically, only model sibtex-5 was able to increase performance for the infNDCG@10 results with respect to the baseline. The linear combination of results sibtex-3 and sibtex-4 to obtain sibtex-5 improved a -0.3% and -8.7% decrease in performance for sibtex-3 and sibtex-4 models, respectively, to a positive outcome of +9.32% for the infNDCG@10 metric. Similarly, for the P@10 (+partial) metric, i.e., considering partial answers as relevant, only model sibtex-5 outperforms the baseline. Again, surprisingly, the linear combination of two negative performances (or neutral, in this case for sibtex-3) with respect to the baseline resulted in an increase in performance of +6.6% for the P@10 (+partial) metric. Finally, all the models outperformed the baseline for the P@10 (-partial) metric, i.e., considering partial answers as not relevant, with emphasis on the model sibtex-5, which improved the baseline by +8.3%. Despite the high difference in relative performance between the models, there is not statistically significant difference between them.

Table 6 presents the overall bioCADDIE challenge performance results aggregated over the participant's top systems for each metric in addition to our individual best results. In particular, the model sibtex-5 achieved the highest inferred average precision among the 10 participants. If we take the overall median result as baseline, our best system would improve the baseline by +22.4% for the infAP metric. Our results for the infNDCG and P@10 (+partial) metrics are not as expressive, for which there is only a marginal improvement over the median score of +0.2% and +0.4%, respectively. For these two metrics, our best system was ranked into 5 out of 10. For the NDCG@10 and P@10 (-partial) metrics, our best system was ranked 3/10 and 2/10, and improved the overall median score by +8.0% and +11.8%, respectively. To be able to compare the overall systems, we computed the Unanimous Improvement Ratio (UIR) metric [38] using the participant's best results achieve in the official run. While this metric is meaningless from the information retrieval viewpoint, it is useful to understand how our overall strategies behave with respect to the other participants. According to this metric, our system ranks overall on top 2 with a URI score of 0.51.

Finally, using the results of all participants, we computed the correlation matrix for the official metrics. The correlation coefficients are displayed in Table 7. Metrics NDCG@10, P@10 (+partial) and P@10 (-partial) present among them moderately strong to strong correlations ($0.6 < \rho \leq 1.0$) while the other metrics show modest to moderate correlations ($0.2 < \rho \leq 0.6$). Particularly, there is a high correlation ($\rho = 0.86$) between the P@10 (+partial) and NDCG@10 metrics, which can be indeed verified on our system's results (see Table 5). Nonetheless, while our system ranks at top 5 for the P@10 (+partial) metric, it ranks at top 3 for the NDCG@10 metric.

## *Post-official assessment*

Due to the small size of the original training dataset (6 queries), it was a challenge to tune all model parameters for the official run. Hence, as described in the Methods section many of them were empirically set. With the disclosure of the challenge results and the gold standard for the test queries, we performed a new experiment where we split the gold standard results into 5-folds, each fold containing 12 training and 3 test queries, and assess the models using cross-validation (see Table S3 in the supplementary materials for parameter setting values). We consider only the results of the test queries because their quality is very different from the training set, due to the pooling method performed after the participants submission. Additionally, it allows us to compare the new results with the official results. The results obtained are displayed in Table 8. First, we can notice a considerable improvement of the baseline model in comparison with the official results (infAP: +20.0%; infNDCG: +7.0%; and P@10 (+partial): +17.9%). Indeed, the performance of the models based only the DFR algorithm have on average improvements of +18.7% and +9.2% for the infAP and infNDCG metrics, respectively. On the other hand, the performance of the model based only on the RTRL similarity measure (i.e., sibtex-4) was more robust for these metrics, varying only -0.03% and +0.29%, respectively. Differently from the official results, the best infAP and infNDCG outcomes are now achieved using sibtex-3 and sibtex-2 models, respectively. Sibtex-5 is still the highest performer for the P@10 (+partial) metric. However, sibtex-4 and sibtex-5 models no longer outperform the baseline for the infAP and infNDCG metrics. Furthermore, in this new setting, the increase in performance of models sibtex-2 and sibtex-3 for the metric infAP is statistically significant (*P*=.049 and *P*=.022).

Table 8 also shows the results for the models sibtex-2, sibtex-3 and sibtex-4 without the post-processing phase (result categorization). We can see that a significant categorization gain is verified only for metric P@10 (+partial) of the sibtex-4 model (+1.3%). For the other metrics and models, the change in performance is marginal. Thus, the actual gain in performance for models sibtex-2 and sibtex-3 in comparison with the baseline derives from the query expansion in the pre-processing phase.

## *Query-wise analyses*

To understand how the different models perform at the query level, we computed the individual query infAP, infNDCG and P@10 (+partial) metrics as showed in Figure 2, Figure 3 and Figure 4, respectively. We use the results of the cross validated experiment since we believe they are better tuned compared to the official run. For infAP metric (Figure 2), we can notice that best performing model, sibtex-3, outperforms all the other models for 4 queries (T1, T6, T8 and T14) while worst performing model, sibtex-1, outperforms all other models for only one query (T3). Models sibtex-2, sibtex-4 and sibtex-5 completely outperform the other models for 3 queries each. For infNDCG metric (Figure 3), model sibtex-1 and sibtex-4 completely outperform the other models for 5 queries each while model sibtex-3, the worst performer for this metric, does not outperform all the other models for any query (conversely, note that on average sibtex-3 has the second best infNDCG). Lastly, for P@10 (+partial) metric (Figure 4), model sibtex-4 outperforms the other models for queries T1, T2, T7 and T9 while models sibtex-2 and sibtex-3 do not outperform all the other models for any query. These results are opposed to the average P@10 (+partial) marks, for which sibtex-3 and sibtex-2 models are, together with sibtex-5, the best performers and sibtex-4 has the poorest results. Thus, from the individual queries view for the tested set, the enhancing characteristics of the models seem to be query specific, without a particular dominant feature that improves the overall results. Indeed, only query T9, *Search for data of all types related to the ob gene in obese M. musculus across all databases*, has consistent highest performance for model sibtex-3, outperforming the other models for infAP, infNDCG and P@10 (+partial) metrics. On the other side, query T1, *Find protein sequencing data related to bacterial chemotaxis across all databases*, has highest performance marks for different models (sibtex-3 for infAP; sibtex-1 for infNDCG; sibtex-4 for P@10 (+partial)). Indeed, query T1 shows peculiar results. While it has high precision for the top 10 retrieved datasets (Figure 4), it has the lowest recall among the 15 official queries for the 5 models (mean recall=0.25; SD=0.01), likely leading to the low infAP (Figure 2). Moreover, for infNDCG metric (Figure 3), it presents average (RTRL model) to high (DFR models) marks depending on the base similarity model.

At the query level, we can also note that queries T3, T4 and T15 obtain consistently good results, having their performance marks higher than the 3$^{rd}$ quartile threshold for the different metrics for at least one model. On the other hand, queries T5 and T8 have overall worst performances, being below the 1$^{st}$ quartile threshold for all the models for at least two metrics. In the example of T5, *Search for gene expression and genetic deletion data that mention CD69 in memory augmentation studies across all databases*, the key relevant terms are *CD69* and *memory*. However, they are not present simultaneously in any of the datasets judged as relevant. Additionally, as showed in Figure 2 and Figure 3, for query T5 the expansion step did not succeed in bringing additional equivalent terms that could enrich the original query. Indeed, it is the combination of the DFR and RTLR similarity models that boosts relevant datasets to the top ranks as showed in Figure 4 (note that models sibtex-2 and sibtex-3 are the low performers). On the other hand, for query T8, *Search for proteomic data related to regulation of calcium in blind D. melanogaster*, the query expansion process is able to effectively enrich the query, resulting in a 17% higher recall (from 42 to 49) for models sibtex-2 and sibtex-3 in comparison with sibtex-1. Nevertheless, as showed in Figure 2, the overall results for query T8 is still poor.

## *Query expansion*

In this section, we analyse the effect of the different collections for query expansion on the information retrieval performance. Table 9 shows some examples of terms expanded using the bioCADDIE, PMC and Medline corpora. The terms were expanded using the 10 nearest vectors in the embedding space. The similarity score between the original term and the expanded term is also provided. As we can notice, the terms expanded can be syntactic variations of the original term, e.g., *cancer* and *cancers*; semantic synonyms as in *cancer* and *tumour*; term subclasses, e.g., *cancer* and *carcinoma*; term superclasses, e.g., *human* and *vertebrate*; but also, just a simple co-occurrence term that has neither syntactic nor semantic relations to the original term, e.g., *human* and *also* and *repair* and *mus7*. In

Table 10, we present the performance results of the information retrieval system using the different collections in the pre-processing phase to expand the terms. Using the bioCADDIE collection for training the word vectors marginally improves the baseline (sibtex-1) for the infNDCG metric while it degrades the results for the other metrics. On the other hand, using the PMC collection marginally improves the infAP metric while also degrading the other two metrics. Finally, the Medline collection improves all the baseline metrics, increasing the infAP metric by +4.1%, the infNDCG metric by +3.4%, and the P@10 (+partial) metric by +3.4%.

# Discussion

In this manuscript, we described our methods to improve ranking of biomedical datasets using an information retrieval engine. The biomedical community is increasingly developing data management systems based on advanced text analytics to cope with extremely large size and variety of datasets involved in scientific research [6-10,39]. Indeed, this trend is not specific to publicly available data but also directly affects large pharma and biotech companies, which are struggling to transform silo-based drug development models into more integrated platforms powered with advanced text analytics [40]. Here the dataset retrieval model is especially adequate if we consider that user queries in such large companies are likely to involve simultaneously a large set of orthogonal entities (a chemical compound, a therapeutic indication, a genetic profile, …) from a large set of perspective (efficacy, toxicity, mode of action…) and modalities (structured laboratory results, sequences, chemical structures, textual reports, images…).

Our work was developed in the context of the 2016 bioCADDIE Dataset Retrieval Challenge, which aimed at developing innovative methods to retrieve biomedical datasets relevant to researcher's needs. As strategies to enhance retrieval performance, our system proposes a query expansion model based on word embeddings, a similarity model (RTRL) that discriminates key-relevant terms from the user query, and a query and dataset categorizer that boosts the ranking of matching query-dataset results. Our retrieval system was evaluated using a corpus of 800k datasets and 15 official assessment queries. It achieved the highest inferred average precision among the 10 challenge participants. If the aggregation of the best official metrics per participant is taken into account, it also achieved very competitive results, figuring on an overall top-2 position. Nevertheless, due to the limited size of the assessment set, i.e., 15 queries, it is hard to extract statistically significant comparisons among different systems and methods.

The approach developed in this manuscript is data driven rather than model driven. As such, it could with some effort be transposed to other domains. In particular, its application could be interesting in the health big data context. One of the main challenges for health and clinical research in large and distributed environments is to index and search for existing datasets, in particular for cohort identification in privacy-preserving frameworks. There are many systems being investigated and developed that tackles the issue of preserving individual and patient privacy and confidentiality so that personal datasets can be ethically shared within trusted networks [41,42,43]. While these systems are very good in keeping data safe and private, they usually lack functionalities that would allow the value of healthcare datasets to be unlocked. If inserted in such secure frameworks, the methodology described here could be a first step for indexing and effectively searching for shareable datasets to enable, for example, posterior patient counting and individual consent requests. A main requirement for employing our approach to related use-cases would be the availability of a corpus composed of dataset metadata extracts pointing to encoded dataset object identifiers, which would roughly describe the content of the dataset, as provided for example by the European Genome-phenome Archive [43].

## *Query expansion approach*

The analysis of the results show that use of the query expansion algorithm based on continuous embedding space enhanced the average performance of the ranking system. Exploiting a Medline collection with 200k abstracts as the input for the word2vec algorithm resulted in a query expansion model that provides statistically significant improvements in comparison with the baseline (sibtex-3: +5.0% for the infAP metric). Given the diversity and richness of bioCADDIE dataset, we argue that it is impractical and unnecessary to implement syntactic and semantic query expansion services using the myriad of biomedical information resources scattered in the web if the goal is to produce richer user queries. Instead, we propose to use either the bioCADDIE dataset itself or a comprehensive biomedical corpus resource, such as Medline or PMC, fed to a continuous embedding space algorithm to extract syntactically and semantically closest concepts to a query term. This approach has two advantages: i) instead of having a query expansion service connected to external resources that are prone to changes over time, we need only a single resource that can be stored and managed locally; ii) it might be able to provide more up-to-date terms, since new terms are likely to appear first in the scientific literature than on curated terminologies. The main drawback of such approach is that, as showed in Table 9, there is little control on the type of terms resulted from the expansion, i.e., the algorithm can produce synonyms, syntactic variations, super classes, sub classes, etc. Thus, this approach should fit more with a user supervised expansion than as deployed in our system.

One of the main challenges for query expansion is to define which expanded terms are actually relevant for the querying task. Syntactically and semantically close terms, as those provided by word2vec, are natural candidates. However, the degree of proximity can vary based on several factors, such as the type of query, the input corpus, the expansion model, etc. In our setup, to select and weight the candidate terms from the embedding vector space, we combine a term neighbourhood (trained for k=10 in the official run) with a similarity based threshold, by taking into account the cosine similarity of the expanded term as a query weight variable. There are many other possibilities for query expansion though, such as training a word similarity cut-off, creating word clusters in the embedding space (e.g., based on k-means) or even performing the expansion per query rather than per term. During the training phase, we have indeed assessed a different method, based purely on the term similarity. In our experiments, the neighbourhood based method described in the manuscript yielded better performance (e.g. for sibtex-2: sim=0.80 -> infAP=0.0505; sim=0.85 -> infAP=0.0530; sim=0.90 -> infAP=0.0542; sim=0.95 -> infAP=0.0556; k=10 -> infAP=0.0570). Nevertheless, we were more interested at exploring different biomedical corpora as expansion sources and consider the assessment of the various query expansion threshold algorithms as a subject for another work.

### *RTRL similarity measure algorithm*

Despite the simplicity of our similarity model, it provided competitive results in comparison with more complex and powerful models, such as the DFR framework. Specifically, our model seemed to be more robust to different short query and training scenarios. While there was a large performance variance between the official run and the cross-validated run for the DFR-based system (infAP: 18.3%; infNDCG: 8.6%), the RTRL similarity measure model showed a small performance difference under these distinctive tuning scenarios (infAP: 2.9%; infNDCG: 0.6%). Indeed, using Eq. 5 and Eq. 6 to represent *if-then* fuzzy rule-based forms and Eq. 7 to represent *if-then* fuzzy rule-based form with multiple conjunctive antecedents, our similarity measure approach can be reduced to a fuzzy logic based method. As has been showed by Gupta et al. [44], these systems outperform ranking models, such as BM25-Okapi, in certain ranking scenarios. One of main assumptions of the RTRL similarity measure algorithm is that the system shall deal with small user queries, i.e., with 10 or less terms. Therefore, logic components, such as query term frequency, can be removed from the equation as the number of word occurrences in the query is unlikely to be relevant in these cases (most likely one occurrence per query). In addition, the algorithm implements only a concise set of key information retrieval ranking rules. Thus, even compared to other fuzzy logic models, our approach is still simpler. We believe that the robustness of the results may derive from this characteristic, as the system might be less prone to overfitting. However, a larger training and assessment sets are still necessary to generalize such observations.

### *Query and dataset categorization*

Our strategy for boosting performance using result classification according to the query constraints did not achieve significant improvements for most of the metrics and systems. As we can see from Table 8, apart from the +1.3% increase in the P@10 (+partial) metric for model sibtex-4, all the other models and metrics have marginally significant performance increase or decrease when post-processing classification is included in the pipeline. This could be due to the fact that the similarity measure model is already correctly capturing the query constraints and there is little room for improvement. It could be also due to the failure of our categorisation system to capture correctly the query constraint, the result dataset class or both. First, the extraction of the query constraints is performed using an ad-hoc algorithm that searches for keyword combinations in the query and match them against the UniProt classification definition. Second, since our categorization algorithm was validated using scientific abstracts [29], it may be behaving differently for the bioCADDIE datasets. We have no gold standard to explicitly validate both classification approaches and the implicitly validation used via changes in the information retrieval performance did not show significant results. However, we believe that a larger assessment set together with more investigation on the classifier tuning are still needed to draw final conclusions about this approach.

### *Train and test set drifting*

Looking at the results of the training and official assessment phases, we notice a large variation between the model's performances. As expected, the official judgement, which was initially obtained following the same methodology of the training set, was enriched by pooling the participant's methods. Indeed, while the training set contains on average 49 relevant (including partially) datasets judged per query, the official set has on average 259, i.e., a 5-fold increase. We believe that the enrichment of the assessment set, with new complementary datasets to those retrieved by the official engines, resulted in a significant performance shift between the training and official phases by better capturing the relevance of the results provided by our models. This performance drifting is seeing on both intra- and inter-model comparisons. When we make a model-wise performance comparison between training (Table 4) and official (Table 5) phases, on average, our models achieved during the official phase superior performance levels of 520%, 61% and 692% for the infAP, infNDCG and P@10 (+partial) metrics, respectively. Also, for inter-model changes, the performance of the sibtex-4 model in the infAP metric, for example, degraded by -28.6% compared to the baseline model in the training phase (Table 4). For this same metric, the sibtex-4 model outperformed the baseline by +15.0% in the official results (Table 5). Thus, it seems that the quality of the training set was too poor to allow realistic parameter tuning. Furthermore, it seems that our original similarity measure model, RTRL, was able to bring new relevant documents to the top ranking, which were not initially captured by the 4 information retrieval engines used to generate the official gold standard (training and test). These new relevant documents were then deemed as relevant in the post-judgement phase. To overcome the issues with the training set, we performed a re-tuning of the system's parameters using the official assessment gold standard in a 5-fold cross-validation fashion. With the enhanced gold standard, we managed to better optimize the model parameters, in particular, for the systems using the DFR algorithm, for which the term frequency normalization changed from 1 in the training phase to 33 in the cross-validation assessment (see *Training parameters* section - Table S2 and Table S3 - in the supplementary materials). Indeed, we think that the size and depth of the training/test set are the main limitations of the results.

## *Assessment methodology*

An important point to notice is that the official gold standard results contain unjudged datasets (weight -1) obtained during the pooling phase, in addition to the relevant (weight 2), partially relevant (weight 1) and not relevant (weight 0) judgments. Having unjudged results in the gold standard is not an issue per se, as long as this factor is properly propagated to the evaluation tool, in particular, when unjudged datasets compose a significant part of the relevance judgement, as is the case of bioCADDIE's official results (86% of the 142805 total results are unjudged datasets). Nevertheless, it seems not to be case of the official trec_eval assessment tool provided by the challenge organisers. We made few experiments with our official results, where we i) changed the weight of the unjudged results to 0, i.e., we deem them as not relevant, and ii) removed the unjudged results from the relevance file. In both cases, there were average improvements of +6% for the infAP metric and of +47% for the infNDCG metric. Therefore, the unjudged results have a higher negative impact on the official results than results judged as not relevant, a highly unexpected observation. This issue shall be also verified in the other participants results and shall be taken into account in future comparisons using bioCADDIE's benchmark.

## *Online deployment*

Our system was designed to work on real time both for data ingestion and for querying. For indexing, we use Terrier, which allows incremental document indexing. Nevertheless, we are not tightly coupled to an information retrieval engine. We could change easily to ElasticSearch for example. For the similarity model, the main parameter is the query length (number of terms). Thus, we expect it to provide robust performance in different scenarios without the need of constant fine-tuning, as it was showed for the official and cross-validated results. After the challenge, we have implemented the document classifier and query expansion modules as software as a service for Novartis (NIBR TX TMS). The document classification service takes on average 30ms to classify a document on a non-dedicated server with 40 Intel® Xeon® CPU E5-2690 v2 @ 3.00GHz and 757GB of memory. In the challenge, we post-processed 50% of the top returned datasets. If we fix this number to the top 100 retrieved results, it will take 3s per user query. For the query classification step, the time is negligible since we perform only string matching (skip-gram, bigram, unigram). Finally, for the query expansion service, it takes on average 68ms to expand a term (online phase). In the challenge, each query had on average 5.6 terms (excluding stopwords). Hence, it should take only 0.4s to expand an average user query.

## Conclusion

Data management systems are increasingly being developed to integrate, store and provide scientists with easy access to health and life science research data. In this work, we introduce a ranking pipeline to improve search of biomedical research datasets. Our system was assessed in the context of the 2016 bioCADDIE Dataset Retrieval Challenge. Our approach achieved competitive results in comparison with the challenge participants, improving by +22.3% the median inferred average precision of the overall participant's best submissions. Based on these preliminary results, we believe that the dataset retrieval engine solution proposed in this work can be an alternative to research data management frameworks that need enhanced dataset search. Nevertheless, further experiments with larger assessment sets shall still be performed to achieve more conclusive results.

## Funding

These experiments were supported by Novartis (NIBR NX TMS).

*Conflict of interest*. None declared.

# Figures and Tables

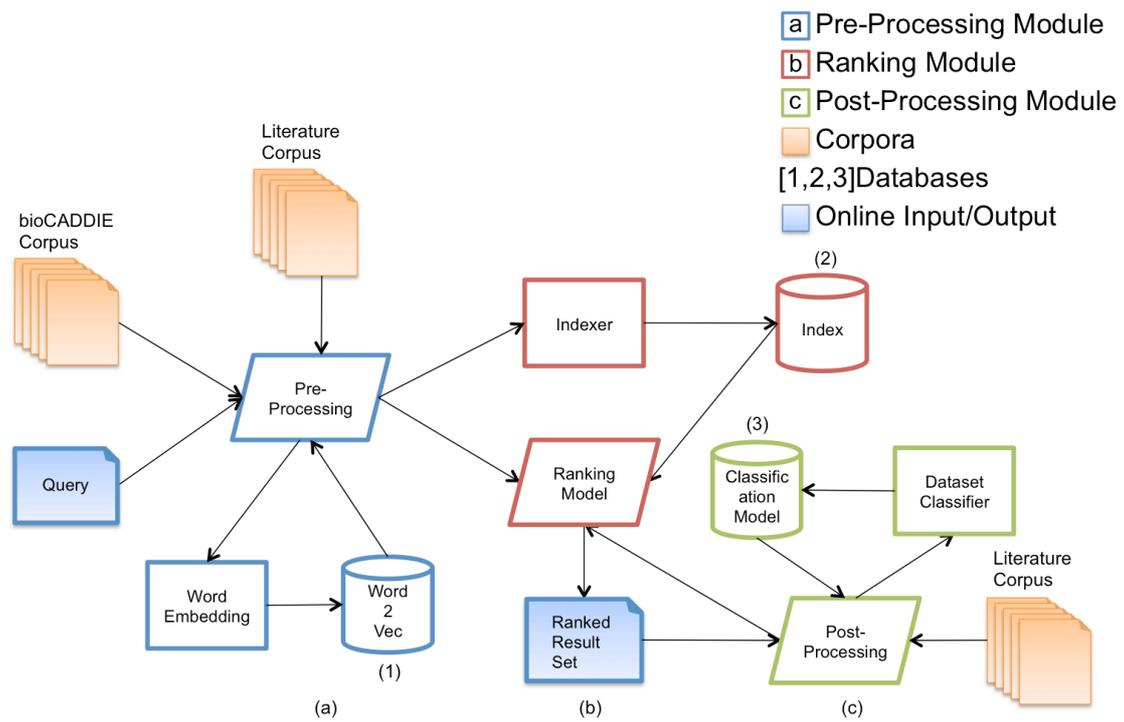

Figure 1 – Architecture of the SIB Text Mining dataset retrieval system.

Table 1 – Example of a bioCADDIE dataset.

```
<DOC>
 <DOCNO>215676</DOCNO>
 <TITLE>VGlut-F-800286</TITLE>
 <REPOSITORY>neuromorpho_030116</REPOSITORY>
 <METADATA>
{
  "dataItem": {
     "dataTypes": ["dataset", "organism", "anatomicalPart", "treatment", "cell", "studyGroup", "dimension", "dataRepository", "organization"]
   },
  "studyGroup": {
     "name": "Control"
   },
  "anatomicalPart": {
     "name": ["Left Antennal Lobe", "Not reported"]
   },
  "dataRepository": {
     "abbreviation": "NeuroMorpho",
     "homePage": "http://neuromorpho.org/",
     "name": "NeuroMorpho.Org",
     "ID": "SCR:002145"
   },
  "dataset": {
     "downloadURL": "http://neuromorpho.org/neuron_info.jsp?neuron_name=VGlut-F-800286",
     "note": "Cell types and Brain regions were assigned with a <a href=\"techDocFlyData.jsp?code=1\">heuristic process</a> based on available metadata. This dataset was processed with a <a href=\"techDocFlyData.jsp?code=2\">streamlined automated variant</a> of the standardization procedure, additional details of which are published <a href=\http://www.ncbi.nlm.nih.gov/pubmed/?term=25576225\ target=\"_blank\">here</a>. Digital reconstruction used a <a href=\"http://www.ncbi.nlm.nih.gov/pubmed/?term=23028271\" target=\"_blank\">custom method</a> after image segmentation by Amira.",
     "ID": "27187",
     "title": "VGlut-F-800286"
   },
  "cell": {
     "name": ["Principal cell", "Glutamatergic neuron", "day8 Born"]
   },
  "treatment": {
     "title": "Green fluorescent protein (GFP)"
   },
  "organization": {
     "abbreviation": "GMU",
     "homePage": "http://www.gmu.edu/",
     "name": "George Mason University",
     "ID": "SCR:011213"
   },
  "organism": {
     "strain": "VGlut-Gal4",
     "scientificName": "",
     "name": "Drosophila melanogaster",
     "gender": "Female"
   },
  "dimension": [{"name": "age"}, {"name": "weight"}, {"name": "soma surface area"}, {"name": "surface area"}, {"name": "volume"}]
}
</METADATA>
</DOC>
```

Table 2 – Repositories and the dataset distribution used in the 2016 bioCADDIE Dataset Retrieval Challenge.

| Repository | Datasets | | Avg dataset size | Avg number of attributes |
|---|---|---|---|---|
| | (#) | (%) | (KB) | (#) |
| **ClinicalTrials** | 192500 | 24.257 | 4.0 | 45 |
| **BioProject** | 155850 | 19.638 | 1.1 | 11 |
| **PDB** | 113493 | 14.301 | 4.0 | 147 |
| **GEO** | 105033 | 13.235 | 0.4 | 14 |
| **Dryad** | 67455 | 8.500 | 2.1 | 38 |
| **ArrayExpress** | 60881 | 7.672 | 1.6 | 12 |
| **Dataverse** | 60303 | 7.599 | 1.9 | 20 |
| **NeuroMorpho** | 34082 | 4.295 | 1.3 | 38 |
| **Gemma** | 2285 | 0.288 | 1.6 | 9 |
| **ProteomeXchange** | 1716 | 0.216 | 1.1 | 32 |
| **PhenDisco** | 429 | 0.054 | 67.2 | 36 |
| **NursaDatasets** | 389 | 0.049 | 1.6 | 34 |
| **MPD** | 235 | 0.030 | 2.2 | 36 |
| **PeptideAtlas** | 76 | 0.010 | 3.2 | 24 |
| **PhysioBank** | 70 | 0.009 | 1.2 | 18 |
| **CIA** | 63 | 0.008 | 1.0 | 32 |
| **CTN** | 46 | 0.006 | 1.4 | 17 |
| **OpenfMRI** | 36 | 0.005 | 1.5 | 20 |
| **CVRG** | 29 | 0.004 | 2.0 | 20 |
| **YPED** | 21 | 0.003 | 1.7 | 25 |

Table 3 – Example of user queries used in the bioCADDIE challenge.

> i) *Search for data on neural brain tissue in transgenic mice related to Huntington's disease*
> ii) *Search for gene expression datasets on photo transduction and regulation of calcium in blind D. melanogaster*
> iii) *Find data of all types on the regulation of DNA repair related to the estrogen signaling pathway in breast cancer patients across all databases*
> iv) *Search for protein aggregation and gene expression data regarding aging across all databases*

Table 4 – Performance results obtained using the training queries.

| Model | infAP | infNDCG | P@10 (+partial) |
|---|---|---|---|
| sibtex-1 | 0.0570 | 0.2714 | 0.0833 |
| sibtex-2 | 0.0573 | 0.2717 | 0.0833 |
| sibtex-3 | 0.0575 | 0.2724 | 0.0833 |
| sibtex-4 | 0.0407 | 0.2016 | 0.1000 |
| sibtex-5 | 0.0539 | 0.2520 | 0.1000 |

Table 5 - Official performance results for the SIB Text Mining models.

| Model | infAP | infNDCG | P@10 (+partial) | NDCG@10 | P@10 (-partial) |
|---|---|---|---|---|---|
| sibtex-1 | 0.3006 | 0.3898 | 0.7067 | 0.5736 | 0.3200 |
| sibtex-2 | 0.2997 | 0.3864 | 0.7067 | 0.5726 | 0.3267 |
| sibtex-3 | 0.3008 | 0.3875 | 0.7067 | 0.5718 | 0.3267 |
| sibtex-4 | 0.3458 | 0.4258 | 0.6600 | 0.5237 | 0.3267 |
| sibtex-5 | 0.3664 | 0.4188 | 0.7533 | 0.6271 | 0.3467 |

Table 6 – bioCADDIE official results: SIB Text Mining individual best score and relative rank, and aggregated participant's best score stats. UIR: Unanimous Improvement Ratio.

| | Stats | infAP | infNDCG | P@10 (+partial) | NDCG@10 | P@10 (-partial) | UIR |
|---|---|---|---|---|---|---|---|
| SIB Text Mining | rank | 1/10 | 5/10 | 5/10 | 3/10 | 2/10 | 2/10 |
| | score | 0.3664 | 0.4258 | 0.7533 | 0.6271 | 0.3467 | 0.51 |
| All participants | median | 0.2994 | 0.4250 | 0.7500 | 0.5806 | 0.3100 | 0.13 |
| | min | 0.0876 | 0.3580 | 0.5333 | 0.4265 | 0.1600 | -1.00 |
| | 1st quartile | 0.2570 | 0.3954 | 0.7150 | 0.5546 | 0.2700 | -0.43 |
| | 3rd quartile | 0.3219 | 0.4433 | 0.7600 | 0.6234 | 0.3333 | 0.40 |
| | max | 0.3664 | 0.5132 | 0.8267 | 0.6861 | 0.4267 | 0.82 |

Table 7 – Metrics correlation matrix – Kendall method.

| | infAP | infNDCG | P@10 (+partial) | NDCG@10 | P@10 (-partial) |
|---|---|---|---|---|---|
| **infAP** | 1 | | | | |
| **infNDCG** | 0.50 | 1 | | | |
| **P@10 (+partial)** | 0.43 | 0.28 | 1 | | |
| **NDCG@10** | 0.42 | 0.29 | 0.86 | 1 | |
| **P@10 (-partial)** | 0.60 | 0.39 | 0.63 | 0.69 | 1 |

Table 8 - Performance results obtained using 5-fold cross-validation with the post-judgment gold standard. *Results skipping original post-processing phase.

| Model | infAP | infNDCG | P@10 (+partial) |
|---|---|---|---|
| sibtex-1 | 0.3557 | 0.4235 | 0.7327 |
| sibtex-2* | 0.3704 | 0.4377 | 0.7511 |
| sibtex-2 | 0.3704 | 0.4378 | 0.7511 |
| sibtex-3* | 0.3735 | 0.4367 | 0.7544 |
| sibtex-3 | 0.3734 | 0.4365 | 0.7544 |
| sibtex-4* | 0.3454 | 0.4216 | 0.7067 |
| sibtex-4 | 0.3441 | 0.4197 | 0.7156 |
| sibtex-5 | 0.3514 | 0.4199 | 0.7578 |

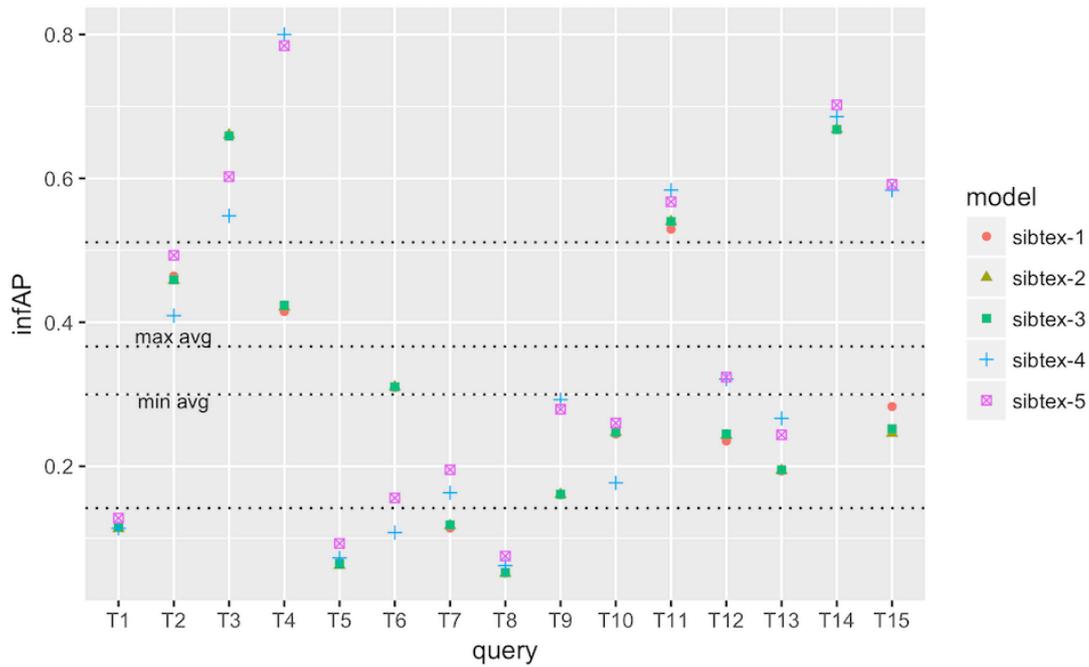

Figure 2 – infAP performance at the query level for SIB Text Mining models. Results obtained using 5-fold cross-validation. Lowest horizontal line: 1st quartile computed for all results. Highest horizontal line: 3rd quartile computed for all results. *min avg* horizontal line: minimum infAP among the 5 models. *max avg* horizontal line: maximum infAP among the 5 models.

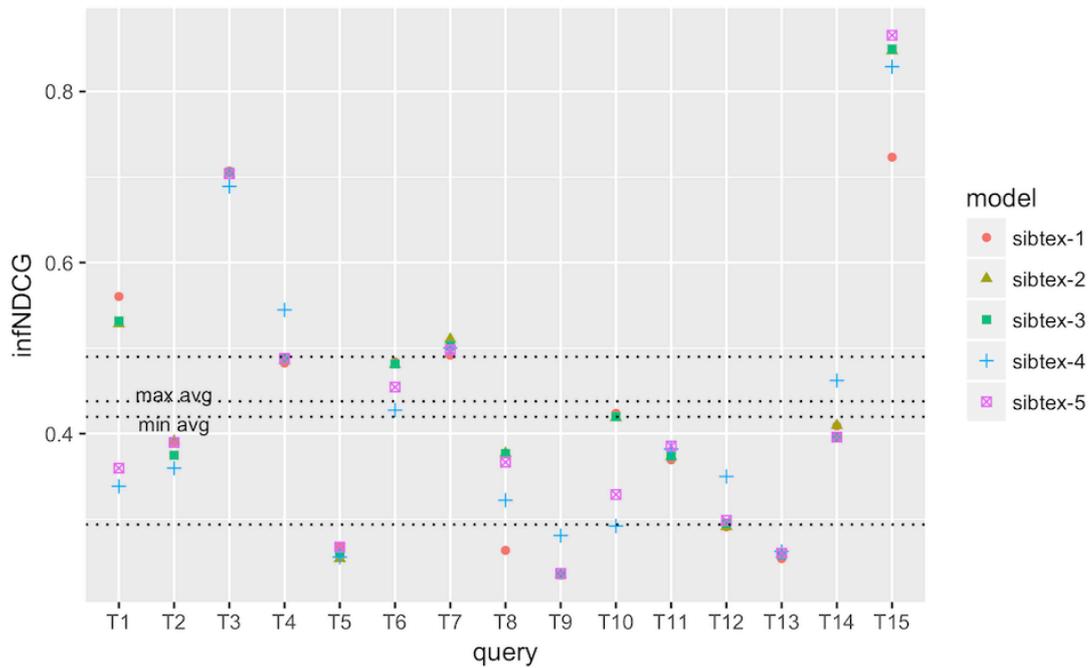

Figure 3 – infNDCG performance at the query level for SIB Text Mining models. Results obtained using 5-fold cross-validation. Lowest horizontal line: 1st quartile computed for all results. Highest horizontal line: 3rd quartile computed for all results. *min avg* horizontal line: minimum infNDCG among the 5 models. *max avg* horizontal line: maximum infNDCG among the 5 models.

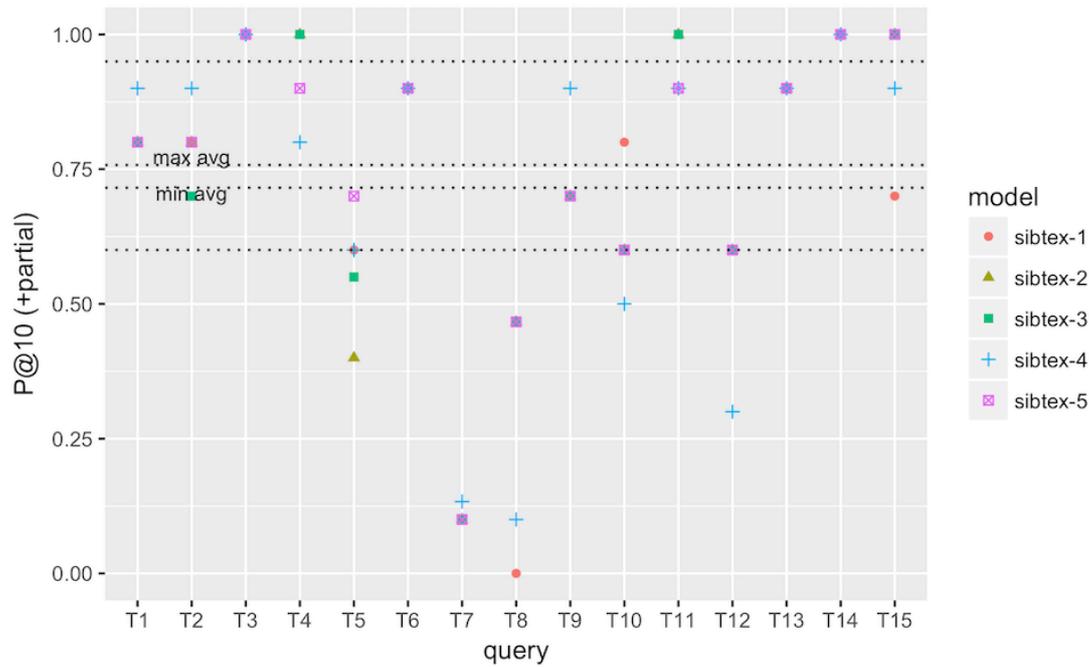

Figure 4 – P@10 (+partial) performance at the query level for SIB Text Mining models. Results obtained using 5-fold cross-validation. Lowest horizontal line: 1st quartile computed for all results. Highest horizontal line: 3rd quartile computed for all results. *min avg* horizontal line: minimum P@10 (+partial) among the 5 models. *max avg* horizontal line: maximum P@10 (+partial) among the 5 models.

Table 9 - Example of term expansion with embedding vectors trained on different corpora.

| Term | bioCADDIE | | PMC | | Medline | |
|---|---|---|---|---|---|---|
| | expansion | score | expansion | score | expansion | score |
| cancer | carcinoma | 0.737 | carcinoma | 0.674 | breast | 0.889 |
| | cancers | 0.720 | tumor | 0.616 | cancers | 0.855 |
| | adenocarcinoma | 0.669 | cancers | 0.585 | prostate | 0.801 |
| | malignancies | 0.626 | tumour | 0.583 | colorectal | 0.794 |
| | lymphoma | 0.621 | glioma | 0.559 | tumor | 0.779 |
| | tumor | 0.616 | carcinomas | 0.545 | carcinoma | 0.773 |
| | transplantation | 0.613 | melanoma | 0.534 | tumorigenesis | 0.761 |
| | transplant | 0.570 | tumorigenesis | 0.532 | tumors | 0.754 |
| | myeloma | 0.563 | carcinogenesis | 0.530 | metastasis | 0.745 |
| | carcinogenesis | 0.559 | glioblastoma | 0.520 | tumour | 0.733 |
| human | bovine | 0.553 | mammalian | 0.582 | mouse | 0.756 |
| | porcine | 0.542 | murine | 0.441 | mammalian | 0.711 |
| | murine | 0.526 | rat | 0.428 | also | 0.661 |
| | mouse | 0.518 | vertebrate | 0.417 | humans | 0.661 |
| | humans | 0.486 | preeclamptic | 0.400 | murine | 0.656 |
| | mammalian | 0.483 | pompe | 0.396 | present | 0.650 |
| | rat | 0.483 | hepa1 | 0.392 | report | 0.644 |
| | chicken | 0.471 | chicken | 0.388 | function | 0.641 |
| | tissue | 0.456 | gpr84 | 0.380 | functional | 0.632 |
| | cellular | 0.456 | cyprinid | 0.373 | well | 0.625 |
| repair | closure | 0.515 | repairthe | 0.597 | damage | 0.794 |
| | metabolism | 0.510 | replication | 0.570 | excision | 0.764 |
| | formation | 0.509 | ssbr | 0.543 | double-strand | 0.727 |
| | grafting | 0.504 | repairing | 0.540 | nucleotide-excision | 0.723 |
| | implantation | 0.502 | damage | 0.516 | damaged | 0.717 |
| | reconstruction | 0.498 | regeneration | 0.499 | breaks | 0.715 |

| | | | | | |
|---|---|---|---|---|---|
| autologous | 0.489 | healing | 0.498 | mus7 | 0.699 |
| testing | 0.474 | detoxification | 0.493 | dsb | 0.697 |
| mobilization | 0.469 | resection | 0.489 | helix-distorting | 0.681 |
| remodeling | 0.468 | processing | 0.479 | post-replication | 0.679 |

Table 10 – Retrieval performance for different collections as query expansion source. The baseline results use no query expansion.

| Collection | infAP | infNDCG | P@10 (+partial) |
|---|---|---|---|
| - (baseline) | 0.3557 | 0.4235 | 0.7267 |
| bioCADDIE | 0.3545 | 0.4243 | 0.7178 |
| PMC | 0.3571 | 0.4216 | 0.7178 |
| Medline | 0.3704 | 0.4377 | 0.7511 |

# Supplementary material

Table S1 – Train queries (EA prefix) and test queries (T prefix).

> EA1 – Find data of all types related to TGF-β signaling pathway across all databases
> EA2 – Find data of all types on synaptic growth and remodeling related to glycolysis in the human brain across all databases
> EA3 – Search for data on BRCA gene mutations and the estrogen signaling pathway in women with stage I breast cancer
> EA4 – Find data of all types on the regulation of DNA repair related to the estrogen signaling pathway in breast cancer patients across all databases
> EA5 – Search for data of all types on multiple sclerosis of all types across all databases
> EA6 – Find data on T-cell homeostasis related to multiple sclerosis across all databases
>
> T1 – Find protein sequencing data related to bacterial chemotaxis across all databases
> T2 – Search for data of all types related to MIP-2 gene related to biliary atresia across all databases
> T3 – Search for all data types related to gene TP53INP1 in relation to p53 activation across all databases
> T4 – Find all data types related to inflammation during oxidative stress in human hepatic cells across all databases
> T5 – Search for gene expression and genetic deletion data that mention CD69 in memory augmentation studies across all databases
> T6 – Search for data of all types related to the LDLR gene related to cardiovascular disease across all databases
> T7 – Search for gene expression datasets on photo transduction and regulation of calcium in blind D. melanogaster
> T8 – Search for proteomic data related to regulation of calcium in blind D. melanogaster
> T9 – Search for data of all types related to the ob gene in obese M. musculus across all databases
> T10 – Search for data of all types related to energy metabolism in obese M. musculus
> T11 – Search for all data for the HTT gene related to Huntington's disease across all databases
> T12 – Search for data on neural brain tissue in transgenic mice related to Huntington's disease
> T13 – Search for all data on the SNCA gene related to Parkinson's disease across all databases
> T14 – Search for data on nerve cells in the substantia nigra in mice across all databases
> T15 – Find data on the NF-κB signaling pathway in MG (Myasthenia gravis) patients

## *Training parameters*

## Official results

Table S2 – Optimized parameters for official submission.

| Model | Term frequency normalization (c) | Query expansion (k) | Classification gain (g) | Non key-relevant term expansion loss (l) | Key-relevant term weight boost (c) | Linear combination coefficient (α) |
|---|---|---|---|---|---|---|
| sibtex-1 | 1.00 | | | | | |
| sibtex-2 | 0.95 | 10 | 0.10 | | | |
| sibtex-3 | 0.95 | 10 | 0.10 | 0.01 | | |
| sibtex-4 | | | 0.10 | | 1 | |
| sibtex-5 | | | | | | 0.5 |

## Post-official cross-validation

Table S3 – Optimized parameters for cross-validation experiment.

| Model | Term | Query | Classificatio | Non key- | Key- | Linear |
|---|---|---|---|---|---|---|

|          | frequency normalization (c) | expansion (k) | n gain (g) | relevant term expansion loss (l) | relevant term weight boost (c) | combination coefficient ($\alpha$) |
|----------|---|---|---|---|---|---|
| sibtex-1 | 33 | | | | | |
| sibtex-2 | 33 | 25 | 0.01 | | | |
| sibtex-3 | 33 | 25 | 0.01 | 0.05 | | |
| sibtex-4 | | | | 0.02 | 1.85 | |
| sibtex-5 | | | | | | 0.7 |